\documentclass[twocolumn,aps,pre,showpacs,floatfix]{revtex4}
\usepackage{amsmath}
\usepackage{amsfonts}
\usepackage{amssymb}
\usepackage{bm}
\usepackage{color}
\usepackage{graphicx}

\newcommand{\msuM}{\frac{m}{M}}
\begin{document}

\title{Macroscopic equations for the adiabatic piston}

\author{Massimo Cencini$^{1,2}$, Luigi Palatella$^{1,3}$, Simone Pigolotti$^{4,5}$ and
Angelo Vulpiani$^{6}$}

\affiliation{
$^1$INFM-CNR, SMC Dipartimento di Fisica, Universit\`a di
Roma ``La Sapienza'', Piazzale A.\ Moro 2, I-00185 Roma, Italy\\
$^2$ISC-CNR Via dei Taurini 19, I-00185 Roma, Italy\\ 
$^3$ Istituto di Scienze dell'Atmosfera e del
Clima del CNR, Str. Provinciale
Lecce-Monteroni km. 1,200, I-73100, Lecce, Italy\\
$^4$Instituto de Fisica Interdisciplinar y Sistemas Complejos IFISC (CSIC-UIB),
  Edificio Mateu Orfila, Campus Universitat Illes Balears, E-07122,
  Palma de Mallorca, Spain\\ 
$^5$ Niels Bohr Institut, Blegdamsvej 17,  DK-2100 Copenhagen, Denmark\\ 
$^6$ Dipartimento di Fisica and INFN
 Universit\`a di Roma ``La Sapienza'', Piazzale A.\ Moro 2, I-00185
 Roma, Italy}

\begin{abstract}
  A simplified version of a classical problem in thermodynamics ---
  the adiabatic piston --- is discussed in the framework of kinetic
  theory. We consider the limit of gases whose relaxation time is
  extremely fast so that the gases contained on the left and right
  chambers of the piston are always in equilibrium (that is the
  molecules are uniformly distributed and their velocities obey the
  Maxwell-Boltzmann distribution) after any collision with the piston.
  Then by using kinetic theory we derive the collision statistics from
  which we obtain a set of ordinary differential equations for the
  evolution of the macroscopic observables (namely the piston average
  velocity and position, the velocity variance and the temperatures of
  the two compartments).  The dynamics of these equations is compared
  with simulations of an ideal gas and a microscopic model of gas
  settled to verify the assumptions used in the derivation.  We show
  that the equations predict an evolution for the macroscopic
  variables which catches the basic features of the problem.  The
  results here presented recover those derived, using a different
  approach, by Gruber, Pache and Lesne in
  \textit{J. Stat. Phys.}~\textbf{108}, 669 (2002) and \textbf{112},
  1177 (2003).

\end{abstract}

\pacs{05.40.-a, 05.70.Ln}

\maketitle

\section{Introduction}
\label{intro}

\begin{figure}[b!]
\centerline{\includegraphics[width=.49\textwidth]{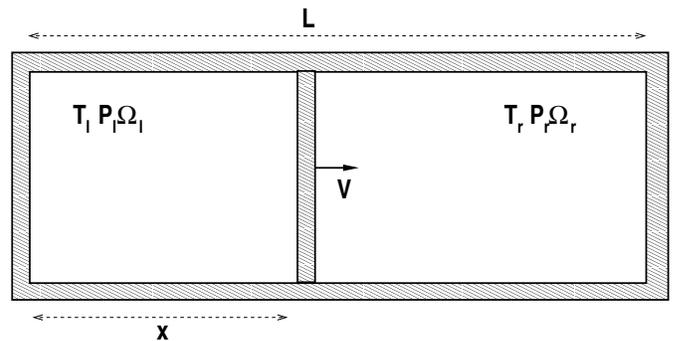}}
\caption{Sketch of the adiabatic piston. The subscripts $l$ and 
$r$ indicate  the left and  right compartments, respectively.}
\label{fig:1}       
\end{figure}

The so-called {\it adiabatic piston} is a long known problem in
classical thermodynamics, which can be stated as
follows~\cite{callen,feynman,gruberlesne}. An isolated cylinder of
length $L$, containing a gas, is divided by an adiabatic wall (no
internal degrees of freedom), \textit{the piston}, into two
compartments (Fig.~\ref{fig:1}).  The initial condition is prepared in
the following way: the piston is kept fixed by a clamp at a given
position $X_0$; the gases in the left ($l$) and right ($r$)
compartments are in equilibrium defined by their pressure, temperature
and volume: $P_{l,r}, T_{l,r}, \Omega_{l,r}$.  By assuming that the
two gases are perfect and composed by $N_{l}=N_r=N$ molecules with
equal masses $m$, the gas state equation $P_{l,r} \Omega_{l,r}=N
T_{l,r}$ holds in both chambers (where the Boltzmann constant is set
to unity by rescaling the temperatures).  Being the piston 
adiabatic, the two subsystems are in equilibrium even if $T_l \ne
T_r$.  At $t\!=\!0$ the clamp is removed and the piston is free to
move without friction with the cylinder.  The nontrivial question is
to predict the system evolution and the final position of the piston
and of the thermodynamic quantities.

At the beginning of last century, the above setup was used as
experimental device for measuring the ratio $c_p/c_v$ of the specific
heat of gases \cite{Ruchardt}, that is linked to the period of the
piston oscillations. Renewed interest on the problem has lead to
recent experiments~\cite{exp1,exp2}.

Meanwhile, several attempts were made to predict the final equilibrium
state by using the laws of thermodynamics only, ending in
controversial answers.  A naive application of the first two laws of
thermodynamics lead to the (wrong) conclusion that the equilibrium
conditions is $P_l / T_l= P_r / T_r$. A more careful
treatment~\cite{curzon} shows that the correct answer is $P_l = P_r$.
However, such a condition says nothing about the final position of the
piston and gas temperatures, which remain undetermined.  Therefore,
equilibrium thermodynamics cannot predict the final state.  To shed
light on the problem one has to cope with the non-equilibrium process
that occurs after the clamp removal.

From a microscopic point of view the adiabatic piston problem for
ideal gases (non-interacting particles) can be described in terms of a
one dimensional model, where the piston is a heavy particle of mass
$M$ much larger that the mass $m$ of the gas molecules, which collide
elastically with the piston. As argued by Feynman, the system first
converges toward a state of mechanical equilibrium with $P_r\approx
P_l$, consistently with the thermodynamic prediction. Then, the
pressure fluctuations, which are asymmetric because $T_l\neq T_r$,
very slowly drive the system toward thermal equilibrium $T_r=T_l$
\cite{feynman}. In this way the final position of the piston and the
thermodynamic quantities are determined.

More recently, the problem was the subject of renewed attention, mainly
stimulated by the talk of Lieb at the Statistical Physics Conference
in 1998 \cite{lieb}, and by the connection of this problem with the
physics of mesoscopic systems \cite{bustamante,crosi1} and
Brownian motors \cite{vandenbroeck1}.

Among the first attempts to understand quantitatively the time
evolution of the adiabatic piston we mention Crosignani et
al.~\cite{crosignani} who introduced a set of ordinary differential
equations for the macroscopic observables. However, this model was
only able to account for the position of the piston in the state of
mechanical equilibrium and not for the final thermodynamic one.

Remaining in the framework of ideal gases, a systematic investigation
in statistical mechanics terms, together with numerical simulations,
has been carried on in the last decade by Gruber and
coworkers~\cite{gruber1,gruber2,gruber3,morrisgruber02,gruberpache02,lesne1,morrisgruber03,lesne}.
In these works the problem has been examined in several limits (see
Ref.~\cite{gruberlesne} for a review). In particular, in the
thermodynamic limit taken by letting the system size $L$ and the
piston mass $M$ to go to infinity by holding fixed the ratios
$\rho_0=N/L$ and $R=mN/M$, it has been shown that the system evolution
can be reduced to a set of ordinary differential equations for the
macroscopic observables (i.e. the gas left/right temperatures, and the
moments of the piston velocity). These equations were obtained by
using the Liouville and Boltzmann equations. Within such an approach
it is possible to control the deviations from Maxwell-Boltzmann
distribution for the gas velocities, observed in the simulations, and
a whole hierarchy of equations can be written for all moments of the
piston velocity. Remarkably, these equations describe not only the
reaching of mechanical equilibrium, which comes from the treatment at
zero order in $\msuM$ \cite{lesne1}, but also the final equilibrium
state, which comes from the first order terms in $\msuM$
\cite{lesne}. These analytical results have been shown by the same
authors to be in agreement with numerical simulations of the ideal gas
piston problem. Though it remains open the problem of a detailed
description of the early stage of the dynamics in which the presence
of shock waves has an important impact on the dynamics. Some recent
attempts in this direction can be found in Ref.~\cite{Mansour}.

When the initial pressures are different, the system phenomenology can
be described as follows~\cite{gruberlesne}.  In a first stage, the
piston oscillates driven by the pressure difference. These
oscillations are then damped till the ``mechanical
equilibrium'' state, $P_r \simeq P_l$ but $T_r \ne T_l$, is reached.
Then, as argued by Feynman, it follows a regime controlled by the
asymmetry in the fluctuations felt by the left/right walls.  This
phase is characterized by a very slow approach to the thermodynamic
equilibrium, $P_r \simeq P_l$ and $T_r \simeq T_l$, with the piston
position fluctuating around the middle of the cylinder.  In the
oscillatory phase, both experiments \cite{exp1,exp2}, numerical
computations and analytical
arguments~\cite{gruberpache02,lesne1,morrisgruber03} have shown the
existence of two different regimes: \textit{weak and strong damping},
the relevant parameter being $R$.  For $R<R_c$ the adiabatic
oscillations of the piston are weakly damped, while for $R>R_c$ they
are over-damped, $R_c$ being $\mathcal{O}(1)$~\cite{lesne1}.

Still in the context of ideal gases, it is worth mentioning some
recent approaches based on dynamical systems theory that have been
developed by Chernov, Lebowitz and Sinai~\cite{sinai}. In this
context, also the case of gases starting from non-equilibrium
conditions has been considered~\cite{lebo1,Caglioti}.  Clearly, the
ultimate goal would be to quantitatively understand the behavior of
the system for an interacting gas, but this seems to be still too
ambitious. Indeed only very few studies
analyzed the case of gas composed by interacting
particles~\cite{vanderbroek2}.

In this paper, we consider a limiting case which has the advantage of
being more tractable while displaying most of the non-trivial features
of the problem. The basic idea of our approach is to assume that the
gases in the two compartments are composed of interacting molecules
and thus characterized by a relaxation time toward the equilibrium
state.  Our main hypothesis is that this time is very short compared
with all the other characteristic times of the system.  In particular,
we require that any fluctuation away from equilibrium (which is
characterized by homogeneously distributed gas molecules with a
Maxwell-Boltzmann velocity statistics) induced by the collision with
the piston is re-adsorbed before the new collision with the piston
walls.  Physically speaking, the efficient re-adsorption of the
fluctuations means that the (mechanical) work done by the piston is
immediately converted into heat; an obvious consequence is that shock
waves are ruled out.  For the sake of simplicity, we also assume that
the gases follow the perfect gas law.  These hypothesis make the
problem tractable while retaining the basic phenomenology of the
original problem.

Although a microscopic model of a gas able to fulfill the above
requirements may sound rather artificial, at a practical level such a
``microscopic model'' can be easily implemented on a computer.  The
basic idea is to start with an equilibrium configuration with
temperatures $T_{l,r}$ for the gases, and then reinitialize the gas
molecules as soon as one particle collides with the piston. The
temperatures are recomputed after the collision and used for
extracting a new configuration of the gas molecules. The procedure is
then repeated.  In the following we shall call such a model
\textit{randomized gas}. Even though, no actual interaction among the
particles is actually considered, one can think that the re-generation
of the gas configuration from an equilibrium one (but with the new
temperature) is the result of such ``unresolved'' interactions.

With the above assumptions for the gas, we will derive a set of
ordinary differential equations for the time evolution of the
macroscopic quantities describing the state of the system. Indeed the
fact that the gas is always homogeneous and following the
Maxwell-Boltzmann distribution allows us to compute the joint
probability density function that, in a given state of the system, the
first colliding gas particle hits the piston in a time $t$ and with a
velocity $v$. Then, by averaging over this joint distribution the
energy and momentum exchange due to the collisions with the piston, we
derive the evolution of the macroscopic observables.  The minimal set
of variables required to have a closed set of equations is made up of
the gas temperatures, the mean piston position, the first and second
moment of the piston velocity. The second moment is required for
accounting the piston fluctuations which, as argued by Feynman, are
crucial for recovering the correct thermodynamic
equilibrium~\cite{feynman,gruber3,lesne}.  The equations are derived
perturbatively up to the first order in $\msuM$. As we will see, though
with a different approach and assumption, these equations are very
similar to those derived by Gruber and coworkers~\cite{lesne,lesne1},
in particular at the zeroth order in $\msuM$ they are identical.

We compare then the evolution of the system obtained by simulations of the
ideal and randomized gas. In particular, the agreement in the first
(mechanical) regime is quantitatively perfect in the case of the randomized
gas. While in the second regime, which is dominated by the fluctuations, the
agreement seems to be only qualitative.  Somehow surprisingly, we found that,
in this regime, a better quantitative agreement seems to be possible
disregarding some $\mathcal{O}(\msuM)$ terms. However, with such terms
excluded the equipartition of energy at equilibrium is violated by the piston.
Some hints to explain these findings could come by higher orders terms in the
expansion. Unfortunately, the computation of the higher order terms is very
cumbersome. Since the newest aspects of our work is in the proposed derivation
and in the introduction of the randomized gas model, we present in this paper
the all approach up to first order in $\msuM$.

The paper is organized as follows. In Sect.~\ref{sec:1} we present our
approach based on the collision statistics and derive the equations
for the macroscopic observables.  In Sect.~\ref{sec:2} we compare the
results of the model with those obtained by simulations.  Discussions
and conclusions can be found in Section~\ref{sec:3}.  In order to
avoid long appendices, the technical material, with the detailed
derivation and all the formulas needed to make explicit the equations,
is presented as \textit{electronic supplementary material} ~\cite{epaps}.

\section{Derivation of the macroscopic equations}
\label{sec:1}

The underlying idea of our approach is to derive a set of
deterministic dynamical equations for the ``macroscopic'' variables
describing the evolution of the thermodynamic state of the system
under the assumption that, at any time, the gases in both chambers are
perfect and at equilibrium. In other words the gases are able to
instantaneously dissipate the fluctuations induced by the collisions
with the piston. Thus a Maxwell-Boltzmann equilibrium state holds
always in both compartments but, in general, with different
temperatures and volumes.

While the above hypothesis define the macroscopic state of the gas, for the
piston the problem is more subtle. We would like to describe its motion on
times longer than the single collisions, that is to average its instantaneous
position and velocity $(X,V)$ over the collisions so to obtain a deterministic
(macroscopic) trajectory defined by the average position $x=\overline{X}$ and
velocity $v_x=\overline{V}$ The symbol $\overline{[\dots]}$ denotes the
average over the collisions. As discussed in the introduction, it is crucial
to account also for the fluctuations of the piston velocity. For this reason,
the second moment of the piston velocity $\overline{V^2}$ is included in the
description. 

In the thermodynamic limit we will consider, one can argue that the
fluctuation of the piston position can be safely ignored. This means that in
the following we will consider the piston position as a deterministic quantity
and we shall use only the mean piston position $x$.

Given the piston position, the gas is characterized by the temperature
$T_{l,r}$, and volume $\Omega_{l,r}$, with $\Omega_l=x$ and
$\Omega_r=L-x$ (we assume a 1d geometry for the sake of
simplicity). Being perfect gases, the pressures are given by the
equation of state $P_{l,r}\Omega_{l,r} = N T_{l,r}$.

In the sequel we show how to derive a set of differential equations
for the evolution of $x$, $v_x$, $\overline{V^2}$ and $T_{l,r}$.  In
order to keep the presentation as simple as possible, here we shall
sketch how the equations can be derived and the averages performed
skipping all the algebra of the computation, which is detailed
in \cite{epaps}.

\subsection{Macroscopic equations from the collision rule}
For the formal derivation of the deterministic equations, we only need
the above discussed assumptions and a microscopic ingredient: the
elastic collision rules
\begin{eqnarray}\label{eqnurto}
V'=V+\frac{2m}{M+m}(v-V)\nonumber\\
v'=v-\frac{2M}{M+m}(v-V)\,.
\end{eqnarray}
Primes denote postcollisional velocities, and $v$
the colliding gas particle velocity.  The quantities we are
interested in are the time derivatives of the macroscopic observables
\begin{eqnarray}
&&\frac{{\rm d}x}{{\rm d}t}=v_x \label{eq:x}\\ 
&&\frac{{\rm d} v_x}{{\rm d}t}=\left\langle
  V'-V\right\rangle \label{eq:vx}\\ 
&&\frac{{\rm d}\overline{V^2}}{{\rm d}t} =\left\langle   V'^2- V^2\right\rangle \label{eq:sigma}\\ &&
\frac{{\rm d}T_{l,r}}{{\rm d}t} = m\left\langle v'^2-v^2\right\rangle_{l,r}
  \,,\label{eq:temp}
\end{eqnarray}
 where we set the Boltzmann constant $k_B=1$.  The time derivatives
 should be computed starting from the collision rules as the averages
 $\langle [\dots]\rangle_{l,r}=
 \overline{[\dots]}_{l,r}/\overline{\delta t}$ suggest,
 $\overline{\delta t}$ being the mean collisions time (for a more
 precise and operative definition see Sect~\ref{subsec:average}). The
 subscripts ${l,r}$ denote averages performed over the collisions with
 particles residing on the left/right compartments.

It is useful to introduce $\sigma^2_V= \overline{V^2}-\overline{V}^2$
which evolves as
\begin{equation}
\frac{{\rm d}\sigma^2_V}{{\rm d}t}= \left\langle
V'^2-V^2\right\rangle-2v_x\frac{{\rm d}v_x}{{\rm d}t}\,,
\label{eq:dsigdt}
\end{equation}
where we used Eqs.~(\ref{eq:vx}) and (\ref{eq:sigma}).  It should be
noted that, at this level, the piston is completely described by 
$v_x$ and $\sigma_V^2$. This amounts to 
assume that its velocity distribution is Gaussian 
\begin{equation}
P(V)= \frac{1}{\sqrt{2\pi \sigma^2_V}} e^{-\frac{(V-v_x)^2}{2\sigma^2_V}}\,.
\end{equation}
Since at the initial time $t=0$, one starts with $v_x=0$ and $\sigma_V^2=0$,
the above probability distribution is initially a $\delta$ function. Plugging
(\ref{eqnurto}) into (\ref{eq:vx}-\ref{eq:temp}), we obtain
\begin{eqnarray}
\frac{{\rm d} v_x}{{\rm d}t}&=& N\frac{2m}{M\!+\!m} (\langle v \rangle \!-\!\langle V \rangle)  \label{eq1:vx}
\\
\frac{{\rm d}\sigma^2_V}{{\rm d}t}&=&N
\frac{4m}{(m\!+\!M)^2} \left[
\!-\!M\langle \sigma^2_V \rangle \!+\! m \langle v^2\rangle\!+\!(M\!-\!m) \langle vV\rangle\!\right.\nonumber \\
&-& \left. (m\!+\!M)v_x\langle v\rangle\!+\!
(m\!+\!M)v_x \langle V \rangle\!-\!M\langle v_x^2 \rangle
\right]
   \label{eq1:sigma}
\\
\frac{{\rm
      d}T_{l}}{{\rm d}t} &=&\frac{4Mm}{(M\!+\!m)^2}\left[M\langle \sigma_V^2\rangle_l+M\langle v_x^2\rangle_l
\right.\nonumber\\
&-&\left.(M\!-\!m)\langle V v \rangle_l -m\langle v^2\rangle_l \right ]
  \,,\label{eq:templ}
\\
\frac{{\rm
      d}T_{r}}{{\rm d}t}&=&\frac{4Mm}{(M\!+\!m)^2}\left[M\langle \sigma_V^2\rangle_r+M\langle v_x^2 \rangle_r\right.\nonumber\\
&-&\left.(M\!-\!m)\langle V v \rangle_r
 -m\langle v^2\rangle_r \right ]
  \,.\label{eq:tempr}
\end{eqnarray}
The equation for $\sigma^2_V$ (\ref{eq1:sigma}) is obtained from
(\ref{eq:dsigdt}) and (\ref{eq:sigma}) by using the collision rules
(\ref{eqnurto}).  In (\ref{eq1:vx}) and (\ref{eq1:sigma}), the
prefactor $N$ appears as a result of a time rescaling, that sets the
time unit to the average collision time, which is order $1/N$. Said
differently, the change of the gas temperatures due to the collision
with the piston is order $1/N$.

Notice that, for reasons that will become clear in the following, the
average of the type $\langle V\,v\rangle$ is different from
$v_x\langle v\rangle$. We anticipate that this difference is
not due to a breakdown of molecular chaos hypothesis (as one may
naively think) but to the fact that the collision statistics depends
on the instantaneous value of the piston velocity. More explicitly,
$v_x\langle v\rangle$ represents only the zeroth order term of
$\langle V\,v\rangle$, and a term coming from the fact that $V$ is a
fluctuating quantity will also appear.

Notice also that the above equations conserve the total (gas plus
piston) energy
\begin{equation}
E=\frac{RM}{m} \frac{T_l+T_r}{2}+ \frac{M}{2}(v_x^2+\sigma^2_V)\,.
\label{eq:energy_conservation}
\end{equation}

The consistent (first order in $\msuM$) equations can then be obtained
from (\ref{eq1:vx}-\ref{eq:tempr}) by expanding the various
prefactors, performing the limit $N\to \infty$ and suitably expanding
around it. Since the procedure is delicate, we proceed step by
step.

\subsection{Thermodynamic limit and formal expansion of the equations}
\label{sec:expandedequations}

First of all we have to specify the limiting procedure, that as
explained in \cite{gruberlesne} can be done in different ways. We are
interested in the limit $N,M,L\to\infty$ in which we keep fixed
$\rho_0={N}/{L}$ and the nondimensional mass ratio $R={Nm}/{M}$.  We
have now to expand around this limit retaining all terms which are
first order in $\msuM$ (and consequently at the first order in
$1/N$). Aiming to make explicit the zeroth and first order terms we
formally write the averages as:
\begin{equation}
\langle [\dots]\rangle=\langle [\dots]\rangle^{(0)}+\langle
[\dots]\rangle^{(1)}\,,
\end{equation} 
where the first and second terms on the r.h.s are the zeroth and first
order terms of the expansion. How to explicitly perform such an
expansion will be explained in the following subsections. We warn the
reader that for maintaining the notation as compact and explicit as
possible in the following we adopt the convection to indicate with
$\langle [\dots]\rangle^{(1)}$ all averages which are
$\mathcal{O}(\msuM)$ irregardless if this comes from the expansion of
the average or from the averaged quantity. For instance, by direct
inspection of Eq.~(\ref{eq1:sigma}) at equilibrium, one easily
realizes that $\sigma^2_V$ is $\mathcal{O}(\msuM)$. Therefore, we shall always
indicate its average with $\langle \sigma^2_V \rangle^{(1)}$. Finally,
notice also that all the terms involving powers of $(V-v_x)$ vanish at
the zeroth order. Keeping in mind these simplifications, the (expanded)
equations become:
\begin{widetext}
\begin{eqnarray}
\strut{\hspace{-.2truecm}}\frac{{\rm d}v_x}{{\rm d}t} \!\!&=& \!\!2R\langle v-v_x \rangle^{(0)}+ 2R \left[ \langle v-v_x
  \rangle^{(1)}-\langle V-v_x \rangle^{(1)}-\msuM \langle v-v_x
  \rangle^{(0)}\right] \label{eq2:vx}
\\
\strut{\hspace{-.1truecm}}\frac{{\rm d}\sigma^2_V}{{\rm d}t}\!\! &=&\! \!4R\left[
-\langle \sigma^2_V \rangle^{(1)} +\msuM \langle (v-v_x)^2  \rangle^{(0)}
+\langle (V-v_x)v \rangle^{(1)}+\langle (V-v_x)v_x \rangle^{(1)}\right]
\label{eq2:sigma}
\\
\strut{\hspace{-.6truecm}}\frac{{\rm d}T_l}{{\rm d}t}\!\!&=& \!\!4m\! \left\{\!
\langle v_x(v_x\!-\!v)\rangle_l^{(0)} \!+\!
\left[\!
\langle v_x(v_x\!-\!v)\rangle_l^{(1)} \!\!\!-\!2\msuM\langle v_x^2 \rangle_l^{(0)} \!\!+\!
3\msuM\langle v_xv\rangle^{(0)}_l \!\!+\!\langle \sigma_V^2\rangle^{(1)}_l 
\right.\right.
-\left.\left.\langle (V\!-\!v_x)v\rangle^{(1)}_l \!\!-\!\msuM\langle v^2\rangle^{(0)}_l
\!\right]\!\right\} \label{eq2:Tl}
\\
\strut{\hspace{-.6truecm}}\frac{{\rm d}T_r}{{\rm d}t}\!\!&=&\!\! 4m\! \left\{\!
\langle v_x(v_x\!-\!v)\rangle_r^{(0)} \!+\!
\left[\!
\langle v_x(v_x\!-\!v)\rangle_r^{(1)} \!\!\!-\!2\msuM\langle v_x^2 \rangle_r^{(0)} \!+\!
3\msuM\langle v_xv\rangle^{(0)}_r \!\!+\!\langle \sigma_V^2\rangle^{(1)}_r 
\right.\right.
-\left.\left.\langle (V\!-\!v_x)v\rangle^{(1)}_r \!\!-\!\msuM\langle v^2\rangle^{(0)}_r
\!\right]\!\right\} \,.\label{eq2:Tr}
\end{eqnarray}
\end{widetext}
Before sketching the way the above averages can be
computed (see Sect.~\ref{subsec:average} and \cite{epaps}) we briefly discuss some properties of  the above equations.

The first observation is that Eqs.~(\ref{eq2:vx}-\ref{eq2:Tr}) ensure
the energy conservation (\ref{eq:energy_conservation}) both at the
zeroth and first order, meaning that the expansion is consistent.
Notice also that the relative importance of the various terms is not
the same at all times.  As the system evolves, their relative weights
change corresponding to the different stages of the evolution,
briefly summarized in the introduction and detailed in the following.
For example, consider $\sigma^2_V$.  At the beginning $\sigma^2_V=0$,
then it grows until it reaches its equilibrium value. Consequently,
the terms which involve the velocity fluctuations are not important at
the beginning. While they become $\mathcal{O}(\msuM)$ and play a crucial
role in the final stage of the system evolution. The opposite is true
for the terms involving the average drift $v_x$, which is close to
zero in the second (Brownian) stage of the evolution, and large in the
first (mechanical) part of the system evolution.  Among the terms
involving the piston velocity fluctuations, we should mention a
special role played by those in which it does not (explicitly) appear
$\sigma^2_V$. These are the terms of the form $\langle
(V-v_x)v_x\rangle^{(1)}_{l,r}$ and $\langle
(V-v_x)v\rangle^{(1)}_{l,r}$, as we will see they will be both
proportional to $\sigma^2_V$ and, as anticipated, find their origin in
the way the fluctuations of $V$ affect the collision statistics (see
next subsections and the supplementary material~\cite{epaps}). However, a closer
inspection shows that $\langle (V-v_x)v_x\rangle^{(1)}_{l,r}$ is very
small at all times. In the first stage the fluctuations are negligible
while in the second stage the average drift is very small. Though we
retained this term in the equations and in the numerical simulations,
one can show that they can be removed without problem. Differently the
terms $\langle (V-v_x)v\rangle^{(1)}_{l,r}$ are very important in the
final stage of the evolution and, as discussed below, for obtaining
the correct equilibrium state.

We are still left with performing the infinite volume limit. We
anticipate here that all the terms appearing in the averages are
proportional to either $1/x$, when coming from a left-chamber average,
or $1/(L-x)$ when coming from a right-chamber average. There is no
other dependence on $x$ or $L$ in equations~(\ref{eq:x}-\ref{eq:temp})
and, consequently, Eqs.~(\ref{eq2:vx}-\ref{eq2:Tr}). This implies that
one can simply introduce the hydrodynamic time $t_H=t/L$ 
and the rescaled $x$ coordinate $x_H=x/L$. With
this rescaling, $L$ does not appear anymore in the equations and there
is no need to perform the limit, meaning that there are no corrections
to the results due to finite size effects. For a simpler comparison
with the simulations, we will keep writing in the following the
equations for finite values of $L$; the corresponding expressions in
the hydrodynamic time can be simply obtained with the above
substitution, that in practice corresponds to set $L=1$.

\subsubsection{Equation at the $0^{th}$ order in $\msuM$: Mechanical Regime}
\label{sec:mecheq}
Let us start a closer inspection of the equations starting from the
zeroth order terms, i.e. assuming $\msuM\to 0$. The velocity
fluctuations of the piston are ignored (meaning the motion of the
piston is purely deterministic) and the final equilibrium position
depends on the initial conditions. The result is anyway nontrivial.
As discussed in~\cite{crosignani,lesne1}, having considered
the dynamics and not only the thermostatics (which only tells us the
equality of pressures), we can now determine the mechanical equilibrium
position. In the sequel, we shall show that at the zeroth order in
$\msuM$ we obtain, by using a different approach, the same equations of
Gruber and coworkers~\cite{lesne1}, and following them we sketch how
the mechanical equilibrium point can be computed.

For obtaining the equations at the $0^{th}$ order, we need to set
$\msuM=0$ and to ignore all averages indicated with the superscript
$^{(1)}$ in (\ref{eq2:vx}), (\ref{eq2:Tl}) and (\ref{eq2:Tr}).  In
other words we only need the averages
\begin{eqnarray}
\langle v \rangle_l^{(0)} &=&\ \frac{T_l}{2x m} \left[1-\mathrm{erf}\left(v_x\sqrt{\frac{m}{2T_l}}\, \right)\right] \label{eq:vmed}\\
\langle v \rangle_r^{(0)} &=& -\frac{T_r}{2(L-x) m} \left[1+\mathrm{erf}\left(v_x\sqrt{\frac{m}{2T_r}}\, \right)\right]\nonumber \,,
\end{eqnarray}
with $\mathrm{erf}(x)=2/\sqrt{\pi} \int_{0}^{x} \mathrm{d}z
\exp(-z^2)$, and the averages 
\begin{eqnarray}
\langle v_x\rangle^{(0)}_{l}\!\!\!\! &=&\!\! \frac{v_x}{x}\!\! \left[\!\sqrt{\frac{T_l}{2\pi m}}e^{-\frac{mv_x^2}{2T_l}}\!-\!
\frac{v_x}{2}\!+\!\frac{v_x}{2}\mathrm{erf}\!\left(v_x\sqrt{\frac{m}{2T_l}}\,\right)\!\right]\label{eq:vxmed}\\
\langle v_x\rangle^{(0)}_{r}\!\!\!\! &=&\!\! \frac{v_x}{L\!-\!x}\!\! \left[\sqrt{\frac{T_r}{2 \pi m }}e^{-\frac{m v_x^2}{2T_r}}\!+\!
\frac{v_x}{2}\!+\!\frac{v_x}{2}\mathrm{erf}\!\left(v_x\sqrt{\frac{m}{2T_r}}\,\right)\!\right]\nonumber\,.
\end{eqnarray}
See the supplements~\cite{epaps} for the derivation of the above expressions.
Assuming $v_x \ll \sqrt{T_{l,r}/m}$ (which is reasonable for
realistic values of the physical parameters), and expanding
(\ref{eq:vmed}) and (\ref{eq:vxmed}) in $v_x$, the equations
(\ref{eq2:vx}) and (\ref{eq2:Tl}-\ref{eq2:Tr}) reads
\begin{eqnarray}
\frac{{\rm d}v_x}{{\rm d}t}&=& \frac{R}{m}\left(
\frac{T_l}{x}-\frac{T_r}{L-x}\right) -\frac{R}{m}\gamma(v_x,T_l,T_r)
v_x\label{eq:vxlin} \\ \frac{{\rm d}T_l}{{\rm
d}t}&=&-\frac{2T_lv_x}{x}
+2\gamma_l(v_x,T_l,T_r)v_x^2\label{eq:Tllin}\\ \frac{{\rm d}T_r}{{\rm
d}t}&=& \frac{2T_r v_x}{(L\!-\!x)}+2\gamma_r(v_x,T_l,T_r)v_x^2
\label{eq:Trlin}
\end{eqnarray}
where for the friction coefficients it holds
$\gamma=\gamma_l+\gamma_r$ ensuring energy conservation
(\ref{eq:energy_conservation}) at the $0^{th}$ order; from
(\ref{eq:vmed}-\ref{eq:vxmed}), at the lowest
order in $v_x$ one has:
\begin{equation}
\gamma_{l,r}(v_x=0,T_l,T_r)= \frac{1}{\Omega_{l,r}}\sqrt{\frac{8mT_{l,r}}{\pi}}\,,
\end{equation}
we remind that $\Omega_l=x$ and $\Omega_r=L-x$.   Note that
  apparently, in the limit $L\to \infty$, $\gamma_{l,r}\to 0$, this is
  not the case if the hydrodynamic rescaling is properly
  applied. Indeed, taking the hydrodynamic limit the above expression
  remains unchanged, keeping in mind that in this case the
  "hydrodynamic volumes" are $\Omega^H_l=x_H$ and
  $\Omega^H_r=1-x_H$. In particular, the damping coefficients go to a
  finite value also in the hydrodynamic limit.  It is worth
remarking, that the linearized equations
(\ref{eq:vxlin}-\ref{eq:Trlin}) coincide with those derived in
Ref.~\cite{lesne1} with a different method.  In the absence of
friction, one can easily see that they describe a purely adiabatic
transformation of a one-dimensional perfect (mono-atomic) gas. Indeed
the first term on the r.h.s. of Eqs.~(\ref{eq:vxlin}) is simply the
pressure difference on the two sides of the piston, while the first
term of Eq.~(\ref{eq:Tllin}) and (\ref{eq:Trlin}) can be obtained by
differentiating with respect to time the equation of an iso-entropic
process, namely
\begin{eqnarray}
T_lx^{c_p/c_v-1}&=&C_l \nonumber\\
T_r (L-x)^{c_p/c_v-1}&=&C_r\,,
\label{eq:adiabatic}
\end{eqnarray}
where $c_p/c_v=3$ is the specific heats ratio and the initial conditions fix
$C_{l,r}=T_{l,r}(0)\Omega_{l,r}^{c_p/c_v-1}(0)$. In the absence of the
friction terms this would give rise to periodic oscillations of the
piston.  As discussed in \cite{crosignani,lesne1}, the friction terms
are responsible for the irreversible evolution toward a state of
mechanical equilibrium for which
\begin{equation}
\frac{N\tilde{T}_l}{\tilde{x}}=P_l=\tilde{P}=P_r=\frac{N\tilde{T}_r}{L-\tilde{x}}\,
\label{eq:equalpress}
\end{equation}
where the tilde indicates the mechanical equilibrium quantities.
Notice that in this framework irreversibility naturally emerges as a
result of the averaging over the collisions~\cite{HOO}.
Eq.~(\ref{eq:equalpress}) is also the result of thermostatics, but it
is not enough to determine $\tilde{x}$ and $\tilde{T}_{l,r}$. Indeed,
the full dynamics given by (\ref{eq:vxlin}-\ref{eq:Trlin}) is needed
to predict such mechanical equilibrium point, as shown in the sequel
where we briefly summarize the results first derived in Gruber et
al.~\cite{lesne1}.  First, notice that (\ref{eq:equalpress}) together
with (\ref{eq:energy_conservation}) tell us that $\tilde{T}_{l,f}=
2T_0\Omega_{l,r}/L$, with $T_0=E/N=(T_l(0)+T_r(0))/2$. So that the
equilibrium pressure is $\tilde{P}=2NT_0/L$.  Second, defining
$Z=\sqrt{T_l}x-\sqrt{T_r}(L-x)$ and by using
(\ref{eq:vxlin}-\ref{eq:Trlin}) one can easily see that ${\rm d}Z/{\rm
d}t=\mathcal{O}(v_x^3)$, which if $v_x\ll 1$ means that $Z$ is
conserved. $Z=const$ provides the missing condition to determine the
mechanical equilibrium. The resulting equation
for the equilibrium point is therefore~\cite{lesne1}:
\begin{equation}
\sqrt{\tilde{T}_l}\tilde{x}-\sqrt{\tilde{T}_r}(L-\tilde{x})=
\sqrt{T_l(0)}x(0)-\sqrt{T_r(0)}(L-x(0))\,,
\label{eq:eqpoint}
\end{equation}
which should be solved for $\tilde{x}$ after plugging $\tilde{T}_l=2T_0
\tilde{x}/L$ and $\tilde{T}_r=2T_0(1-\tilde{x}/L)$.
\subsubsection{Equation at the $1^{th}$ order in $\msuM$: Brownian Regime}
\label{sec:brown}
When the first order (in $\msuM$) terms are retained, the terms in
$\sigma_V^2$ (among which, as discussed above, we have also to
consider the terms $\langle(V-v_x)v\rangle^{(1)}$
and $\langle(V-v_x)v_x\rangle^{(1)}$) allows for energy exchange among the
two compartments mediated by the fluctuation of the piston.  Of
course, such terms start to play a role once the mechanical regime
(described by the $0^{th}$ order terms) is finished, i.e. when the
fluctuations of the piston become relevant. This regime driven by the
fluctuations results from the expansions in $\msuM$ and $1/N$ which, as
it happens commonly in Brownian motor-like systems
\cite{vandenbroeck1}, are intertwined and add new (sometimes
unexpected) features to the dynamics. In particular, in this case one
can show that Eq.~(\ref{eq2:vx}-\ref{eq2:Tr}) evolve toward a
nontrivial stable fixed point corresponding to the thermodynamic
equilibrium, i.e.:
\begin{equation}\label{fixedpoint}
\frac{x_{eq}}{L}=\frac{1}{2}\,,\quad v_x=0\,,\quad T_l=T_r=T_{eq}\,,\quad
{\sigma^2_V}_{eq}=\frac{T_{eq}}{M}\,.
\end{equation}
The last equality cannot be explicitly seen from (\ref{eq2:sigma}),
which simply states that at equilibrium $\langle \sigma^2_V
\rangle^{(1)}=\msuM\langle
v^2\rangle^{(0)}+\langle(V-v_x)v\rangle^{(1)}$. As we discussed,
$\langle(V-v_x)v\rangle^{(1)}\propto \sigma^2_V$ and with the explicit
computation at equilibrium~\cite{epaps} one
can see that ${\sigma^2_V}_{eq}=T_{eq}/M$, which is a pleasant result
since it is in agreement with the condition of equipartition of
energy.  Notice that Eq.~(\ref{fixedpoint}) suggests to interpret
$M\sigma^2_V=T_p$ as the temperature of the piston. 

We conclude this subsection mentioning that the above equations are
similar with the ones obtained by Gruber, Pache and
Lesne~\cite{lesne}. Due to the very long expressions involved in the
equations at the first order, we could not decipher whether they are
exactly equal. At the end of next section we shall discuss the
possible source of differences. However, we stress that in these two
works a different approach and different assumptions were made on the
gases.  In particular, Gruber, Pache and Lesne derived the equations
from an expansion of the Boltzmann and Liouvulle equations.  In some
sense, our and their different assumptions can be seen as two
different ways to close the hierarchy of equations to the second order
and one should expect the phenomenology of the two equations to be,
at least, qualitatively similar.

Section~\ref{sec:2} is devoted to compare the evolution of the
macroscopic observables obtained by integrating
(\ref{eq2:vx}-\ref{eq2:Tr}) with that of the microscopic model.
In the following subsection we detail the procedure by which the averages 
can be computed.

\subsection{Explicitation of the averages}
\label{subsec:average}

In order to finalize our program we have now to make explicit the
averages in (\ref{eq2:vx}-\ref{eq2:Tr}).  Let us start by making
explicit the formal expression of averages such as $\langle
[\dots]\rangle_{l,r}$, which should be interpreted as follows.  Denote
with $G_{l,r}(t,v|V)$ the probability of having a left/right collision
in a time $t$ with a velocity $v$ for the r/l-particle conditioned to
a realization in which the piston has velocity $V$, and indicate with
$G$ their sum, i.e. $G=G_l+G_r$, which is normalized to 1. Then
performing the average $\langle A(v,V,t)\rangle_{l,r}$ of a generic
function $A(v,V,t)$ means:
\begin{equation}
\label{eq:averages}
\langle A(v,V,t)\rangle_{l,r} = \frac{\int  \mathrm{d}t \int \!\!\!\int \mathrm{d}v\mathrm{d}V P(V)
G_{l,r}(t,v|V)\, A(v,V,t)}{\int \mathrm{d}t \int\!\!\!\int \mathrm{d}v \mathrm{d}V\, t\, P(V) G(t,v|V)}\,,
\end{equation}
where $\overline{\delta t}=\int \mathrm{d}t \int\!\!\!\int \mathrm{d}v \mathrm{d}V\, t\, P(V) G(t,v|V)$
is the mean collision time.
\subsubsection{Derivation of  $G_{l,r}(t,v|V)$}
Let us now derive $G_l$ and $G_r$. In particular, we shall compute
them in the thermodynamic limit by explicitly considering the terms
order $1/N$ and consequently $\msuM$. These terms are those entering
the averages we indicated with $\langle [\dots] \rangle^{(1)}_{l,r}$.

We start by the equilibrium distribution of the gases, which is
uniform in the particle positions $y$ and Maxwell-Boltzmann for the
velocities $v$:
\begin{eqnarray}
p_l(y,v)&=&p_l(y)p_l(v)=\frac{1}{x}\frac{1}{\sqrt{2\pi m^{-1}T_l}}
e^{-\frac{mv^2}{2T_l}}\label{eq:equilibrio}\\
p_r(y,v)&=&p_r(y)p_r(v)=\frac{1}{(L-x)}\frac{1}{\sqrt{2\pi m^{-1}T_r}}
e^{-\frac{mv^2}{2T_r}}\,,\nonumber
\end{eqnarray}
that, under our assumptions, describe the gases at all times. Note
that the above distributions depend (parametrically) on the dynamical
variables $x$ and $T_{l,r}$.

From the equilibrium joint distributions (\ref{eq:equilibrio}) we can
derive the probability density $g_{l,r}(t|v,V)$ that a particle on the
left/right collides in a time $t$ given its velocity $v$ and the
macroscopic state of the system defined by the temperatures, the
piston position $x$ and velocity $V$ and the equation of state.  Of
course, most of the weight to such an hitting probability comes from
particles that are close to the piston and that have a large
(negative) relative velocity with it. These particles are far from the
bulk of the gas and in this derivation we assume that we can use the
evolution of free particles to compute their hitting time with the
piston. Within such an assumption, it is easy to realize that the
probability $g_{l,r}(y,v)$ is simply obtained through a change of
variables from (\ref{eq:equilibrio})
\begin{eqnarray}\label{cond}
g_l(t|v,V)\!&=&\!\Theta(t)\Theta\left(\frac{x}{v\!-\!V}\!-\!t\right)\Theta(v\!-\!V)
\frac{v-V}{x}\\&\approx& \nonumber \Theta(t)\Theta(v\!-\!V)
\frac{v-V}{x}\nonumber\\
g_r(t|v,V)\!&=&\!\Theta(t)\Theta\left(\frac{L\!-\!x}{V\!-\!v}\!-\!t\right)\Theta(V\!-\!v)\frac{V\!-\!v}{L\!-\!x}\\&\approx&
\!\Theta(t)\Theta(V\!-\!v)\frac{V\!-\!v}{L\!-\!x}\,,\nonumber\end{eqnarray}
$\Theta$ being the unitary step function. Note that in the second
expression we ignored the second $\Theta$ function. This is justified
by the fact that $t$ indicates the time between two consecutive
collisions, which for $N\gg 1$ is always much shorter than the time
needed for a particle to travel along a whole chamber.
Further, considering only positive
times, the functions $g_{l,r}(t,v|V)$ do not depend on $t$. We shall then
use the compact notation $f_{l,r}(v|V)=g_{l,r}(t,v|V)$.
Notice also that we use the instantaneous piston velocity $V$ and
not its average $v_x$, while we ignore the fluctuations in position.
Moreover, we neglected possible correlations between
the velocities and positions, which amounts to implicitly assume 
molecular chaos.

Given $g_{l,r}(t|v,V)$, the joint probability of having an impact of a
particle with $v$ in a time $t$ is\ :
\begin{eqnarray}
g_{l,r}(t,v|V)=g_{l,r}(t|v,V)p_{l,r}(v)\,.
\end{eqnarray}
The probabilities $F_{l,r}(t_{m})$ that a left/right particle
collides in a time $\leq t_{m}$ is then given by
\begin{eqnarray}
F_{l,r}(t_{m})\!=\!\!\int_0^{t_{m}}\!\! \!\!\!\mathrm{d}t\!\! \int_{-\infty}^\infty \!\!\!\!\!\mathrm{d}v\ 
g_{l,r}(t,v|V)\!=\!t_m\!\! \int_{-\infty}^\infty \!\!\!\!\!\mathrm{d}v
f_{l,r}(v|V)
\label{eq:fmin}
\end{eqnarray}
which we rewrite as
\begin{eqnarray}
F_{l,r}(t_{m})=t_{m}H_{l,r}(x,V,T_{l,r})\,,
\end{eqnarray}
with
\begin{eqnarray}
&&\strut\hspace{-1.truecm} H_l(x,V,T_l)\!=\!\int_{V}^\infty \!\!\!\mathrm{d}v  \frac{v\!-\!V}{\sqrt{2\pi m^{-1}T_l} x}
e^{-\frac{mv^2}{2T_l}}  \label{acca} \\
&&\strut\hspace{-1.0truecm} H_r(x,V,T_r)\!=\!\int_{-\infty}^{V}\! \!\!\!\mathrm{d}v \frac{V\!-\!v}{\sqrt{2\pi m^{-1} T_r} (L\!-\!x)}
e^{-\frac{m v^2}{2T_r}}\,. \nonumber
\end{eqnarray}
In the following we will use the shorthand notation
$H_{l,r}=H_{l,r}(x,V,T_{l,r})$ and
$h_{l,r}=H_{l,r}(x,v_x,T_{l,r})$. 

Considering that we have $N$ particles on both the left and right, the
probability densities that one of them on the left/right impacts the
piston in $x$ with a velocity $v$ are given by
\begin{eqnarray}
&&\strut\hspace{-1.05truecm}G_l(t_{m},\!v)\!=\!N\left[1\!-\!F_l(t_{m})\right]^{N\!-\!1}\!\left[1\!-\!F_r(t_{m})\right]^N\!f_l(v|V)
\nonumber\\
&&\strut\hspace{-1.05truecm}G_r(t_{m},\!v)\!=\!N\!\left[1\!-\!F_r(t_{m})\right]^{N\!-\!1}\!\left[1\!-\!F_l(t_{m})\right]^N \!f_r(v|V).
\label{step1}
\end{eqnarray}
We can now perform the limit $N\!\to\!\infty$ and $t_{m}\!\to\! 0$ holding
$Nt_{m}\!=\!\tau$ fixed. Noticing that $F_{l.r}(t_m) \!\to\! \tau H_{l,r}/N$
and expanding $[1\!-\!F_{l,r}]^{N}\!=\! \exp[N\ln(1\!-\!F_{l,r})]\approx
\exp[N(-F_{l,r}-F_{l.r}^2/2)]$ and retaining terms only up to $1/N$,
we find
\begin{eqnarray}
G_{l,r}(\tau,v|V)\!=\!
\left[1\!+\!\frac{H_{l,r}\tau}{N}\!-\!\frac{H^2_l\!+\!H_r^2}{2N}\tau^2\right]
\!\tilde{G}_{l,r}(\tau,v|V)
\label{step3}
\end{eqnarray}
with
\begin{equation}
\tilde{G}_{l,r}(\tau,v|V)=e^{-\tau(H_l+H_r)} f_{l,r}(v|V)\,.
\end{equation}
By recalling that $N=MR/m$, and aiming to  retain only the
first order $\msuM$ terms (\ref{step3}) can be rewritten as
\begin{eqnarray}
G_{l,r}(\tau,v|V)\!&=&\!
\tilde{G}_{l,r}(\tau,v|V)\!+\! \label{eq:Gfinal}\\
&&\msuM \left(\frac{h_{l,r}\tau}{R}\!-\!\frac{(h^2_l\!+\!h_r^2)}{2R}\tau^2\right)
\tilde{G}_{l,r}(\tau,v|v_x)\nonumber
\end{eqnarray}
where we substituted $h$ in place of $H$ and $v_x$ in place of $V$ in
the part which is already at the first order in $\msuM$. However another
contribution to the ${\mathcal O}(\msuM)$ term comes from the expansion
of the first term in (\ref{eq:Gfinal}).

\subsubsection{Expansion of the average}
We need now to further expand (\ref{eq:Gfinal}) in $V$ around $v_x$,
this can be accomplished by Taylor expanding $\tilde{G}$
\begin{eqnarray}
\tilde{G}(\tau,v|V)&=&\tilde{G}(\tau,v|v_x)+\partial_V\tilde{G}(\tau,v|v_x)
(V-v_x)\nonumber\\ &+&\frac{1}{2}\partial^2_V\tilde{G}(\tau,v|v_x) (V-v_x)^2.
\label{eq:expansion}
\end{eqnarray}
Noticing that $\int \mathrm{d}V P(V) (V-v_x)=0$ and that $\int \mathrm{d}V P(V)
(V-v_x)^2=\sigma^2_V$ we can write, for example, the (expanded) average
collision time:
\begin{eqnarray}
\label{eqtau}
\overline{\delta t}&=&
\frac{1}{h_l+h_r}\left[1-\msuM \frac{h_l^2+h_r^2}{R(h_l+h_r)^2} \right.\\ 
&+&\left.
\frac{h_l+h_r}{2} \sigma^2_V
\int \!\!\!\int \mathrm{d}t\, \mathrm{d}v\, t\, \partial^2_V\tilde{G}(\tau,v|v_x)\right].\nonumber
\end{eqnarray}
\begin{figure*}[t!]
\centerline{\includegraphics[width=1\textwidth]{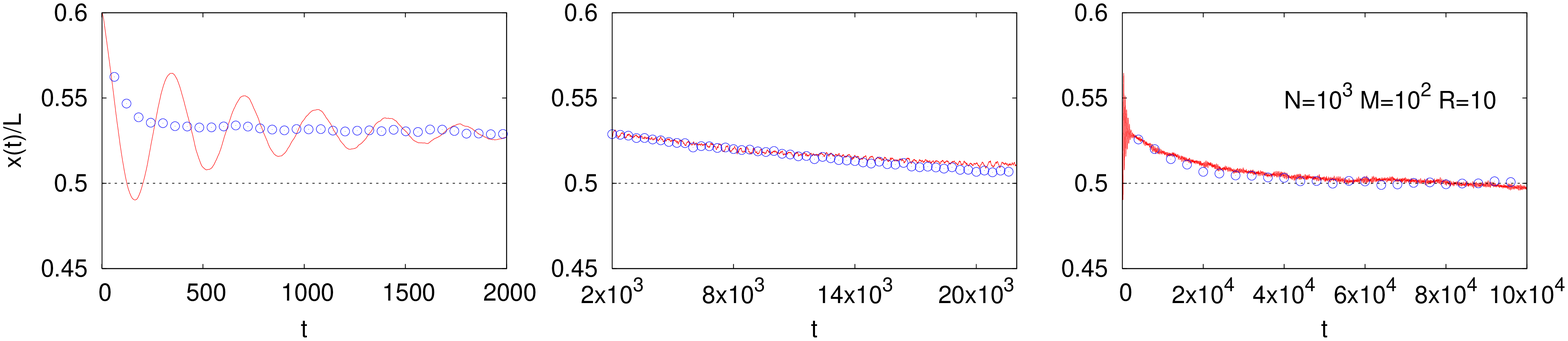}}
\centerline{\includegraphics[width=1\textwidth]{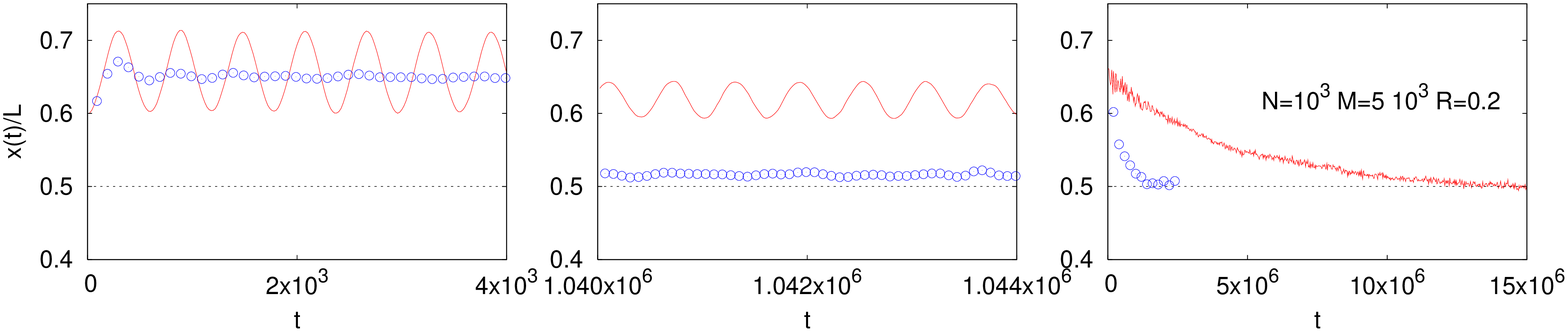}}
\caption{(Color online) (top) Comparison between the evolution of the
 piston position $x(t)$ in a simulation of an ideal gas (red solid
 line) and of the randomized model (blue, open circles). In both
 models we set $N=1000$, $M=100$ and $L=2000$ corresponding to $R=10$,
 at time zero $T_l=40$, $T_r=60$ and $x=0.6L$. To have a cleaned curve
 we performed an average of about $100$ independent realizations.
 (bottom) The of (top) with $N=1000$, $M=5000$ and $L=2000$
 corresponding to $R=0.2$, and the initial state set as $T_l=150$,
 $T_r=50$ and $x=0.6L$.  Here, for the ideal gas, the oscillations
 last for a much longer time and are damped very slowly before the
 Brownian motor like regime sets in; while the randomized gas is much
 more efficient in damping the oscillations.}
\label{fig:evolution}
\end{figure*}

Finally, we can write the correct expansion of
(\ref{eq:averages}) at the first order in $\msuM$, by
plugging all the expanded terms in (\ref{eq:averages})
to have the zero and first order terms of the average of a generic
observable, $\langle A\rangle_{r,l}=\langle
A\rangle^{(0)}_{r,l}+\langle A\rangle^{(1)}_{r,l}$. The result is:
\begin{eqnarray}
\label{eq:recipevera}
&&\strut{\hspace{-.4cm}}\langle A\rangle^{(0)}_{r,l}= \!(h_l\!+\!h_r)
\int\!\!\!\!\int\!\!\!\!\int\! \mathrm{d}\tau \mathrm{d}v \mathrm{d}V\, P(V)\,
\tilde{G}_{l,r}(\tau,\!v|v_x) \, A\\ 
&&\strut{\hspace{-.4cm}}\langle A\rangle^{(1)}_{r,l}=\!\int\!\!\!\!\int\!\!\!\!\int\! \mathrm{d}\tau
\mathrm{d}v \mathrm{d}V\, P(V) \bigg[\partial_V \tilde{G}_{l,r}(\tau,\!v|v_x)
(V\!-\!v_x)\nonumber\\
&&\strut{\hspace{-.4cm}}+\!\frac{1}{2}\partial^{2}_V
\tilde{G}_{l,r}(\tau,\!v|v_x) (V\!-\!v_x)^2\!\bigg]\, A\!+\!\left[\! \msuM
\frac{h_{l,r}}{R(h_l+h_r)}\!-\!\sigma_V^2\right.\nonumber\\
&&\strut{\hspace{-.4cm}} \left.\frac{(h_l+h_r)^2 }{2} \int \!\!\!\!\int\!\!
\mathrm{d}t\, \mathrm{d}v\, t\, \partial^2_V\tilde{G}(\tau,\!v|v_x)\right] \int\!\!\! \mathrm{d}v \,
f_{l,r}(v|v_x) A \label{eq:av1}
\end{eqnarray}
which can be used to compute all the averages in
(\ref{eq2:vx}-\ref{eq2:Tr}).  However we mention that there are
exceptions to the above recipe. For instance, the average of
$A=(V-v_x)^2$, whose result is $\langle\sigma^2_v\rangle^{(1)}$ and
not $\langle\sigma^2_v\rangle^{(0)}$. This is due to the convention
adopted for the average: remember that we chose to write the
superscript $^{(1)}$ also when the $\msuM$ order comes from the
averaged quantity and not from the expansion of the collision
distribution, like in this case.

Notice that in the first term of the r.h.s of (\ref{eq:av1}), as for
the average collision time, we expanded $\tilde{G}$ with the aid of
(\ref{eq:expansion}). Depending on the observable $A$, which may have
or not a linear term in $V-v_x$, also the first derivative of
$\tilde{G}$ may appear.

\section{Comparison between model and molecular dynamics simulations}
\label{sec:2}
In this section we compare the evolution of the macroscopic
observables given by (\ref{eq2:vx}-\ref{eq2:Tr}) with numerical
simulations of the microscopic model. We consider two kinds of
microscopic simulations: the ideal and the randomized gas. The latter
is meant to fulfill the assumptions we made on the gas in deriving 
the equations.  Let us now better clarify how this is realized.

From a computational point of view, it is very easy to realize the
randomized gas, the idea is to let the gases relax in an artificial
manner through a randomization procedure. More precisely, the
simulations are performed in the following way: we generate an
equilibrium configuration of the system, corresponding to given values
of the macroscopic observables. Then the system is let to evolve up to
the first collision of the piston {\em without} interactions among the
gas particles. After the collision, the energy of the gas containing
the colliding particle changes; the fast relaxation of the system is
mimicked by updating the gas temperature (corresponding to the new
energy) and redrawing an equilibrium configuration of the gases
corresponding to the new temperatures (and volumes). Then the process
is iterated. Notice that in this way one keeps track of all the
observables except the piston velocity fluctuation, which needs some
kind of average to be defined/measured. Collisions are evaluated, as
for the ideal gas, with an event driven algorithm.

In Figure~\ref{fig:evolution} we show the evolution of the piston
position by numerical simulations of the ideal and randomized gas for
two different values of $R$ (namely $R=10$ (top) and $R=0.2$ (bottom).
As discussed in the introduction (see \cite{gruberlesne,lesne} for a
more detailed treatment) these two choices correspond to the case of
strongly and weakly damped oscillations, respectively. In both cases
one has two regimes: mechanical and Brownian. The former is
characterized by damped oscillations of the piston, which in the
strongly damped case are very few or nonexistent.  For the ideal gas,
as discussed by Gruber et al. \cite{lesne1,lesne,gruberlesne} the
detailed damping of the oscillations depends on the presence of shock
waves, which are absent in the randomized gas. In fact, the latter is
damped much more efficiently than the former. As shown in the
bottom panel, for $R=0.2$ the evolution is weakly damped and many
oscillations are observable. In this case, the piston is very slow and
the gases perform quasi-adiabatic oscillations, which are damped more
and more mildly as $R\to 0$. Again, also in this case the damping
appear to be much more efficient in the randomized case.
\begin{figure}[t!]
\centerline{\includegraphics[width=0.49\textwidth]{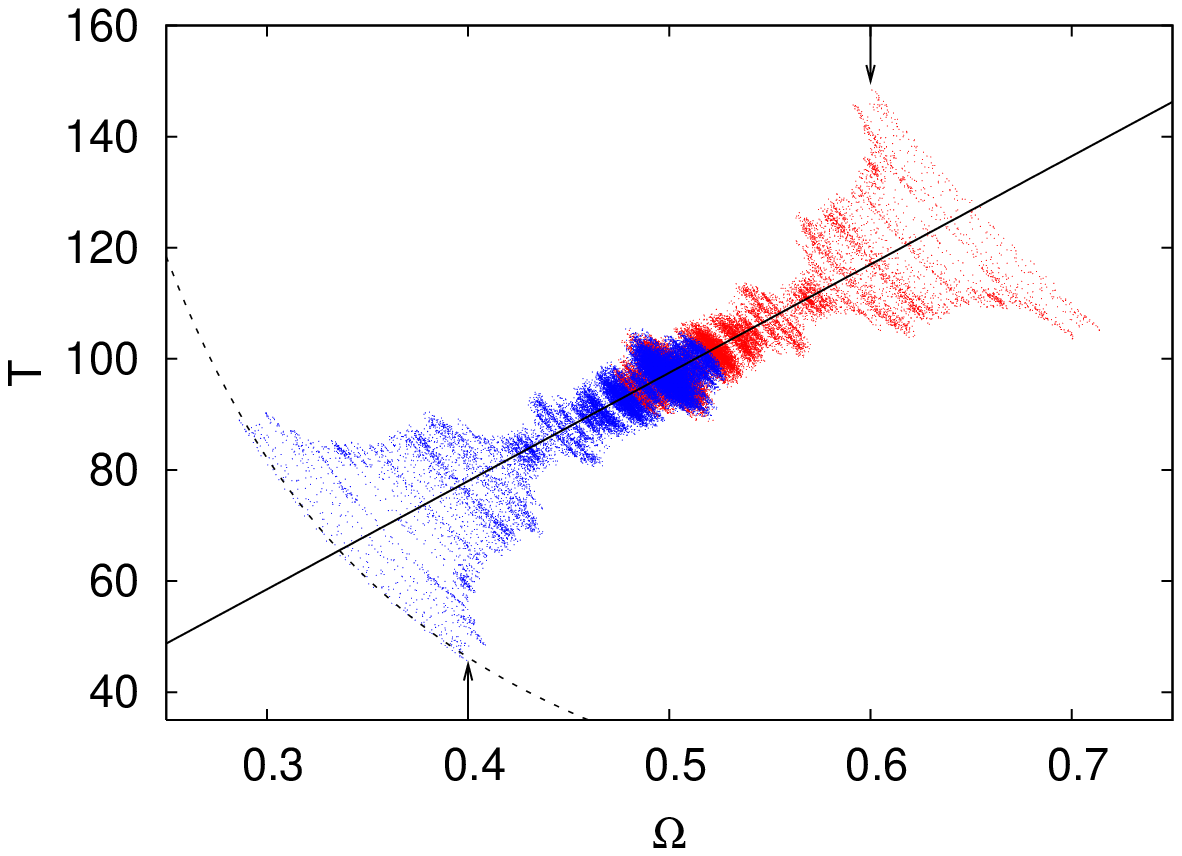}}
\caption{(Color online) Evolution of the system in the temperature
  volume ($T\Omega$) plane.  The dotted line indicates the adiabatic
  phase $T\propto V^{-2}$, while the solid straight line the isobar
  that characterizes the final phase (Brownian motor like regime). The
  rightmost arrow indicates the initial state of the gas in left
  compartment $T_l(0)=150$ $\Omega_l(0)=x(0)=0.6L$ which evolves (red
  dots) toward the final equilibrium in the middle. The leftmost arrow
  indicate the initial state of the gas in left compartment
  $T_l(0)=150$ $\Omega_r(0)=L-x(0)=0.4L$ which evolves (blue dots)
  toward the final equilibrium in the middle.  The simulation is done
  with the ideal gas, for which the initial oscillations are much more
  evident that for the randomized one.  }
\label{fig:TV}   
\end{figure}

The Brownian motor-like regime \cite{vandenbroeck1}, occurs when the
oscillations are completed damped, i.e. the mechanical equilibrium is
realized with approximately equal pressures $P_l\approx P_r$, which
differ only for terms $\mathcal{O}(\msuM)$.  From now on, both the
ideal and randomized gas remain in a state of marginal equilibrium
approximately along the isobar $T_l/x=T_r/(L-x)$, which is the
prediction of thermodynamics (shown in Fig.~\ref{fig:TV} for the ideal
gas only). The Brownian motor-like mechanism is responsible for heat
transfer from the warmer to the colder chamber mediated {\it via} the
piston fluctuations \cite{vandenbroeck1,vanderbroek2}. This stage
occurs on a long timescale (proportional to $M$). For realistic values
of the various parameters, one can realize that the timescales
necessary for reaching the thermodynamic equilibrium state are
enormous. It is therefore very difficult to observe this regime in
experiments, and well controlled numerical simulations are mandatory.
In the final state the two chambers have the same temperatures and the
piston fluctuates reaching equipartition with the same temperature.
It is worth noticing that for the ideal gas, since the molecules
interact only through the collisions with the piston, the probability
distribution of the velocities of the gas molecules may (and actually
do) deviate sensibly from the Maxwell-Boltzmann distribution that is
recovered only after the reaching of the final equilibrium
state~\cite{gruberlesne,lesne}. By construction, this problem is not
present in the randomized gas which is forced to remain
Maxwell-Boltzmann.

In the following we shall compare the evolution of the macroscopic
equations with the simulations.

\subsection{Mechanical regime}
We start the comparison by considering the oscillatory regime in the
weakly damped case. Here we know from previous studies~\cite{lesne1}
that the model is able to correctly predict the period of the
oscillations.

The period of the oscillations in the model can be estimated from the
linearized dynamics at the zeroth order in $\msuM$,
Eqs.~(\ref{eq:vxlin}-\ref{eq:Trlin}).  As discussed in
Sect.~\ref{sec:expandedequations}, the oscillations will occur around
a mechanical equilibrium position $\tilde{x}$ defined by
Eq.~(\ref{eq:eqpoint}). Then to recover the period of the oscillations
it is enough to linearize (\ref{eq:vxlin}-\ref{eq:Trlin}) around the
mechanical equilibrium state defined by $\tilde{x}$, $\tilde{T}_{l,r}$
and $\tilde{v}_x=0$. Linearizing (\ref{eq:Tllin}-\ref{eq:Trlin}) one
can easily recognize that the equation define an iso-entropic process,
i.e. (\ref{eq:adiabatic}), meaning that $T_{l}(x)= \tilde{T}_l(\tilde{x}/x)^{c_p/c_v-1}$ and $T_{r}(x)= \tilde{T}_r[(L-\tilde{x})/(L-x)]^{c_p/c_v-1}$, which
plugged in (\ref{eq:vxlin}) and expanding in $\delta x=x-\tilde{x}$ lead to
the equation of a damped oscillator:
\begin{equation}
\frac{{\rm d}^2\delta x}{{\rm d}t^2}+\frac{R}{m}\gamma \frac{{\rm d}\delta x}{{\rm d}t}+\frac{R}{mN}\frac{c_p}{c_v}\frac{\tilde{P}L}{\tilde{x}(L-\tilde{x})}\delta x=0\,,
\end{equation}
where we recall that $c_p/c_v=3$ and $\tilde{P}=2NT_0/L$ (see
Sec.~\ref{sec:mecheq}). Since the friction coefficient does not
modify the period one immediately gets:
\begin{equation}\label{eq:ruch}
\mathcal{T}=2\pi\sqrt{\frac{Nm\tilde{x}(L-\tilde{x})}{R(c_p/c_v)\tilde{P}L}} \,,
\end{equation}
this formula is at the basis of the measurement of 
the specific heat ratio in experiments~\cite{Ruchardt}.

\begin{figure}[t!]
\centerline{\includegraphics[width=.49\textwidth]{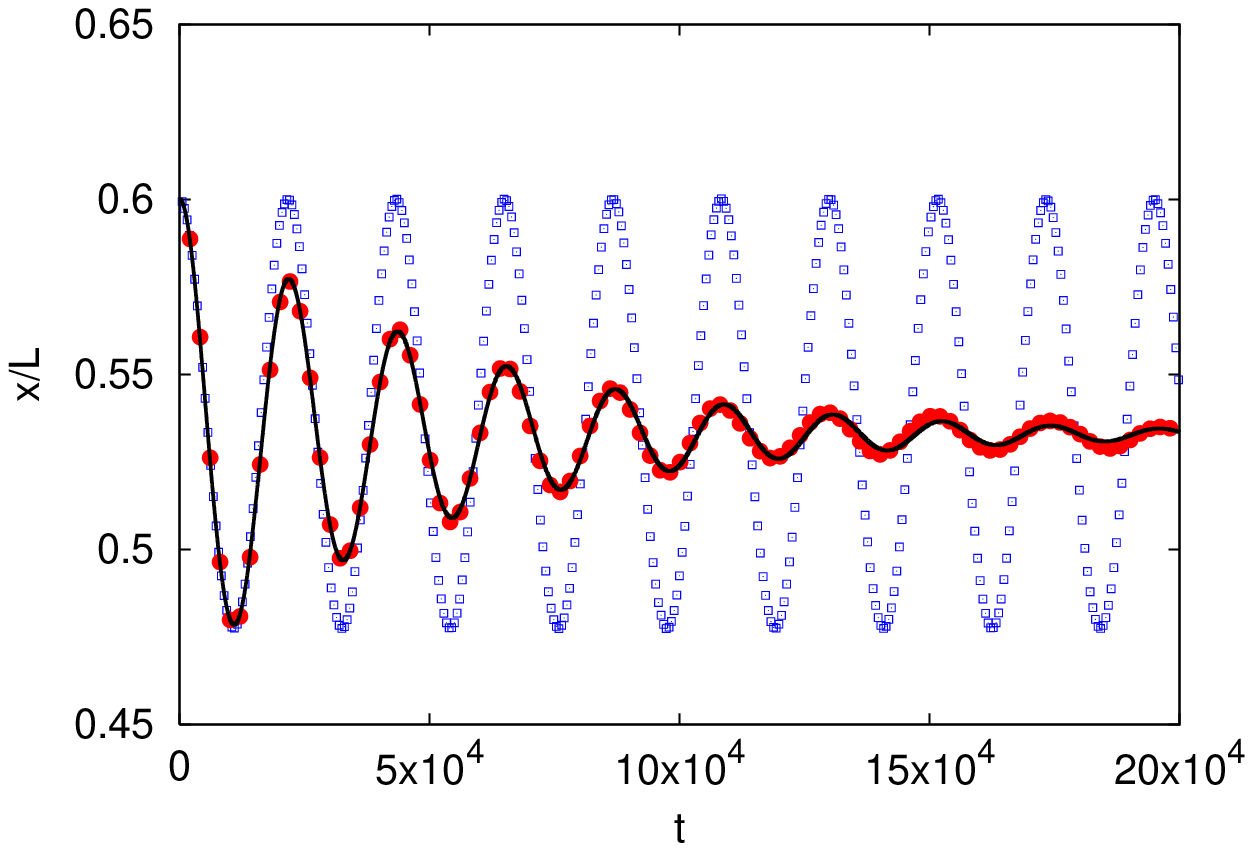}}
\centerline{\includegraphics[width=.49\textwidth]{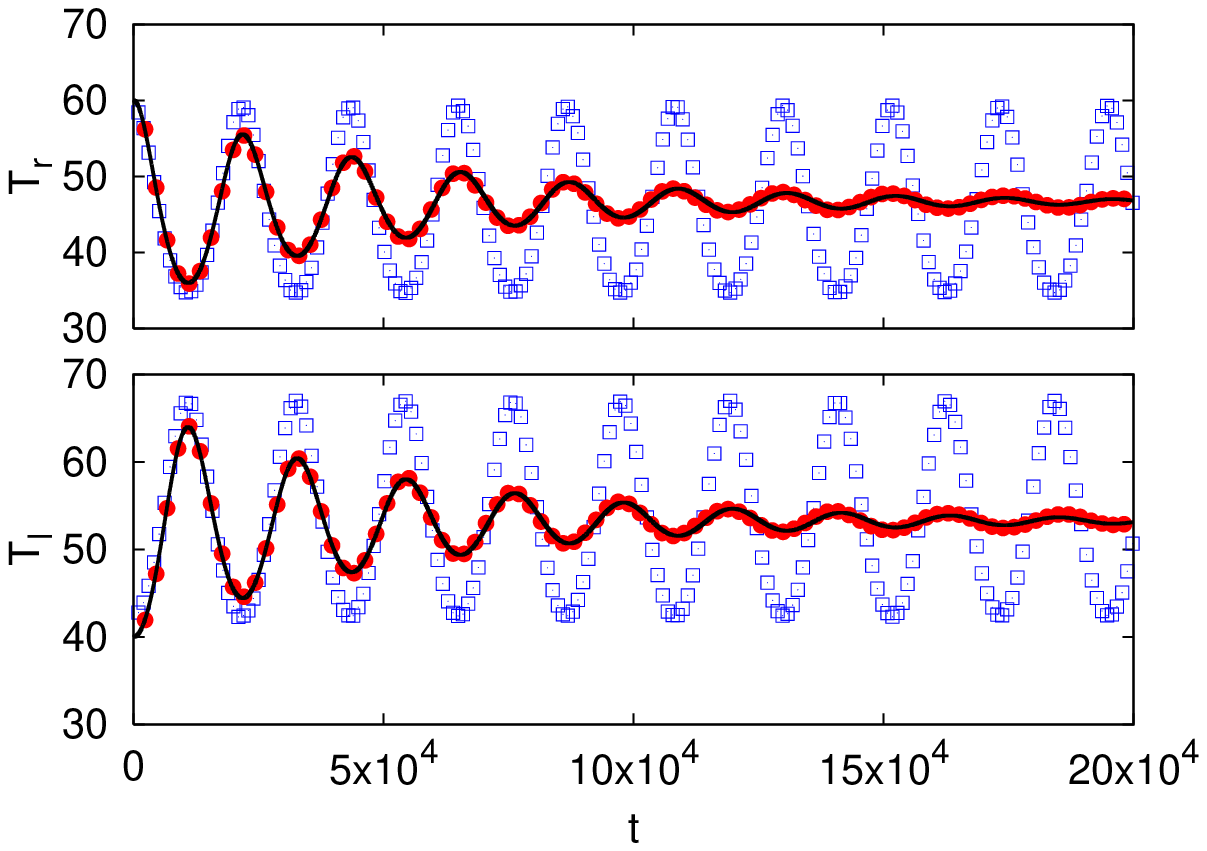}}
\caption{(Color online) Comparison between the simulations of the
  non-interacting 1D gas, the randomized one and the prediction given
  by the macroscopic equations. The parameters are: $T_l=40$,
  $T_r=60$, $L=12000$, $x_0/L=0.6$, $M=10^5$, $N=10^3$ corresponding
  to $R=0.01$. (top) Evolution of the piston position. (bottom)
  Evolution of the temperatures. Blue open squares refer to the ideal
  gas and red filled circles to the randomized gas, the solid line is
  the model (\ref{eq2:vx}-\ref{eq2:Tr}). }
\label{fig:oscill-x}       
\end{figure}

In Fig.~\ref{fig:oscill-x}, we compare the ideal gas simulations with
that of the randomized gas and the numerical integration of the
determinist equations (\ref{eq2:vx}-\ref{eq2:Tr}).  As one can see,
the ideal and randomized gas have the same period, while the damping
is different, and the model is in perfect agreement with the
randomized gas simulations. Integrating the equation at the zeroth
order, we find $\tilde{x}=6.39\times 10^3$ which agrees with the
predicted value (\ref{eq:eqpoint}) and Eq.~(\ref{eq:ruch}) predicts
$\mathcal{T}= 2.17\times 10^4$ which is in very good agreement with
the the measured period in both the randomized and ideal gas.
\begin{figure}[t!]
\centerline{\includegraphics[width=.49\textwidth]{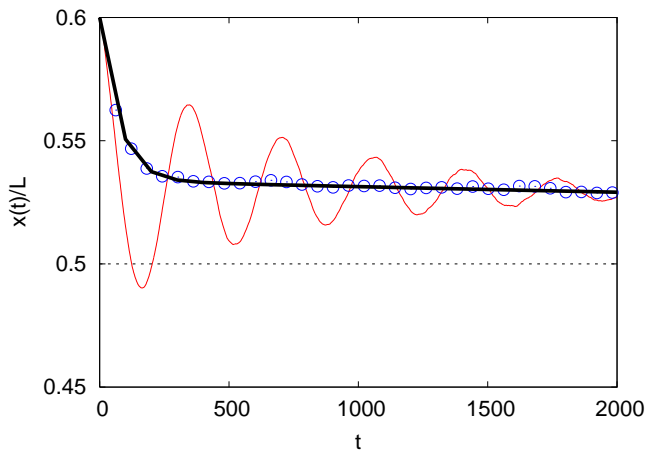}}
\caption{(Color online) The same as the leftmost plot of
  Fig.~\ref{fig:evolution}(top) with superimposed the thick black line
  superimposing on randomized gas data (blue empty circles) is
  obtained by the numerical integration of
  (\ref{eq2:vx}-\ref{eq2:Tr}). The agreement of the model with the
  randomized gas is perfect also for the temperatures (not shown).}
\label{fig:strong}       
\end{figure}

We mention that in Ref.~\cite{lesne1} a detailed study of the period
of the oscillations was reported. Since our linearized equations
coincide with those of Ref.~\cite{lesne1}, we shall not repeat here
this study. It is however interesting to compare the first stage of
the evolution of the ideal and randomized gas with the model in the
case of strong damping. In Fig.~\ref{fig:strong} we show the leftmost
plot of Fig.~\ref{fig:evolution} (top), as one can see although the
model is unable to reproduce the oscillation of the ideal gas, its evolution
coincides with that of the randomized gas.

As discussed in Ref.\cite{gruberlesne}, estimating the decay rate of
the oscillations for the ideal gas is a nontrivial task, since it
requires a detailed study of the dissipation mechanisms of the shock
waves created by the piston motion. These shock waves survive for a
long time in the ideal gas, while they lifetime is expected to be
shorter in the presence of interactions, which should be able to
decrease their coherence.  The randomized gas represents a sort of
limiting case of interaction in which the shock waves are completely
absent. Most likely the absence of shock waves is at the origin of
the faster damping of the oscillations in the randomized gas and of
the very good agreement between its evolution and the one obtained
from the macroscopic equations. It would be interesting to investigate
the transition between weak and strong damping in the randomized
model.  This is far from the aim of the present paper, however the
above results suggest that the critical value of $R$ for the
transition will probably be smaller (but of the same order) of that of
the ideal gas which is order unity.

\subsection{Brownian motor-like regime}

As the mechanical equilibrium is reached $P_l\approx P_r$, the
system slowly evolves, along an approximate isobar, driven by
fluctuations toward the thermodynamical equilibrium defined by
Eq.~(\ref{fixedpoint}), which is the (stable) fixed point of the
ordinary differential equations (\ref{eq2:vx}-\ref{eq2:Tr}).

\begin{figure}[t!]
\centerline{\includegraphics[width=.49\textwidth]{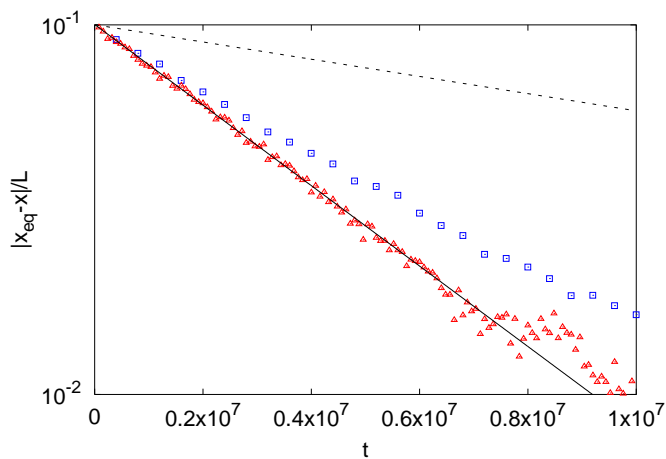}}
\caption{(Color online) Comparison between the molecular dynamics of a
  non-interacting 1D gas and the model in the Brownian motor-like
  regime. Simulations have been performed setting the initial state as
  $T_l=40$, $T_r=60$ and $x(0)=0.4L$ with $M=500$ and $L=12N$ with
  $N=10^4$: red triangles refer to the randomized gas and the blue
  squares to the ideal gas, respectively. The dashed line is the
  prediction of the macroscopic equation, the solid line (which
  perfectly superimposes on the randomized gas data) is explained in
  the text.  The simulation data are obtained by performing an average
  over about $10$ realizations to reduce the fluctuations.}
\label{fig:relax}       
\end{figure}

We shall now compare the evolution given by the macroscopic equations
with that of the ideal and randomized gases in the Brownian regime.
To minimize the possible differences between the ideal and randomized
gases we performed a simulation which starts in a mechanical
equilibrium state having the gas molecules distributed according to
the Maxwell-Boltzmann distribution. For the ideal gas, this might be
not the typical situation: usually when the system arrives to the
mechanical equilibrium from a non-equilibrium state, it may be
strongly non-Maxwellian~\cite{gruberlesne,lesne}.  As exemplified in
Fig.~\ref{fig:relax}, where we show the deviation from the final
equilibrium of the piston position, $|x-x_{eq}|/L$, the relaxation is
exponential. The randomized gas relaxes faster than the ideal one,
likely because the latter develops slightly non-Maxwellian
distribution (we observed that the difference in the relaxation time
tends to diminish as the number of particles is increased). Note that
the simulation has been performed with $R>1$ because the time scale
for reaching the equilibrium is controlled by $M$; with $N=10^4$, as
here, working with $R<1$ would have implied the necessity to reach too
large time scales to study the relaxation.

As shown in the figure, the macroscopic equations (dashed line)
predict a relaxation slower than both the randomized and ideal
gases. This came out as a surprise for us, because we were encouraged
by the very good agreement in the mechanical regime, discussed in the
previous section. With the aim of understanding such a difference we
examined all the terms appearing in the macroscopic equations and we
realized that the mismatch in the relaxation was due to the terms
$\langle (V-v_x)v\rangle_{l,r}^{(1)}$ that, as discussed in
Sect.~\ref{sec:brown}, are those which ensures the equipartition of
energy at equilibrium, i.e. the fact that $M{\sigma_V^2}_{eq}=T_{eq}$.
In particular, eliminating such terms from (\ref{eq2:vx}-\ref{eq2:Tr})
(note that this is not affecting the conservation of energy), we found
a perfect agreement between the macroscopic equations, now modified,
and the randomized gas, as shown by the solid line in the figure.
Nevertheless, with such a modification the energy equipartition is not
verified anymore and at equilibrium $M{\sigma_V^2}_{eq}=2T_{eq}$.

\begin{figure}[t!]
\centerline{\includegraphics[width=.49\textwidth]{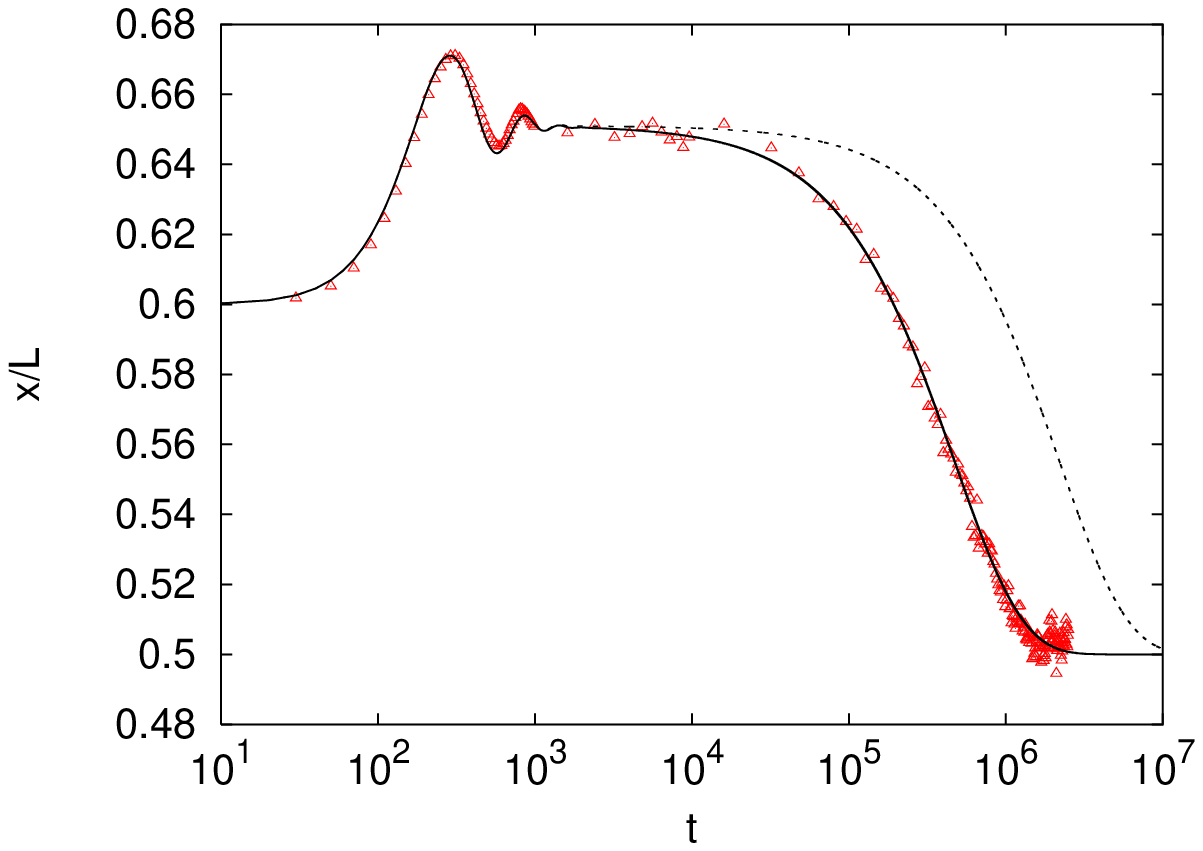}}
\centerline{\includegraphics[width=.49\textwidth]{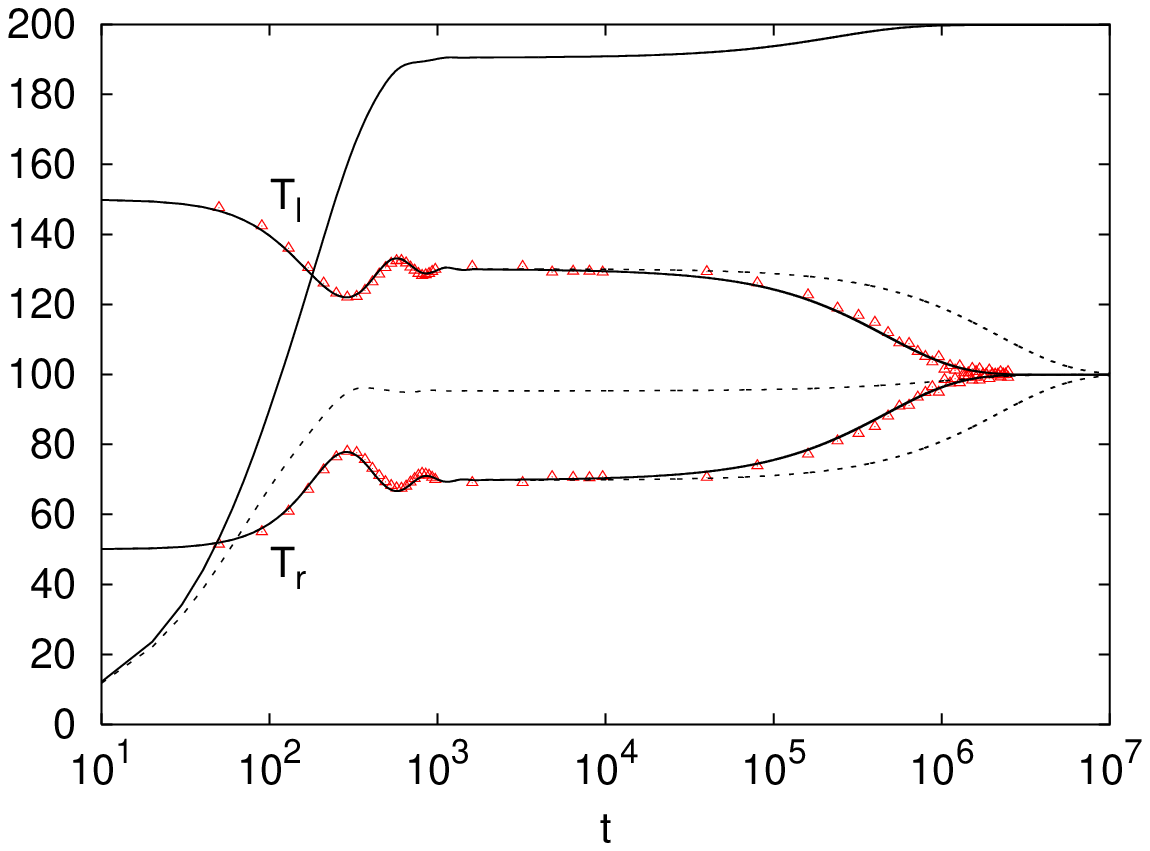}}
\caption{(Color online) (top) Evolution of the piston position for
  parameter as in Fig.~\ref{fig:evolution}(bottom) now plotted in
  log-lin scale. The (red) symbols refer to the randomized gas
  simulations, the dashed curve to the evolution predicted by the
  macroscopic equations and the solid curve (which perfectly
  superimposes on the simulation data) is explained in the
  text. (bottom) Same as (top) but for the evolution of the gas
  temperatures $T_l$ and $T_r$ as in label. The two curved which start
  from zero show the evolution of the piston temperature
  $T_p=M\sigma_V^2$ as obtained from the macroscopic equations (dashed
  curve which saturates to $T_p\approx 100$ and as explained in the
  text the solid one which saturates ad $T_p\approx 200$.}
\label{fig:relax2}       
\end{figure}
One may think that the agreement is incidental, and so we did a more
severe test. In Fig.~\ref{fig:relax2} we show the evolution of the
piston position (top) and of the gas temperatures (bottom) with the
simulations of the randomized gas previously shown in
Fig.~\ref{fig:evolution}(bottom). As one can see, the modified (solid
lines) and original (dashed lines) macroscopic equations generate, a
part from $\sigma^2_V$, two indistinguishable dynamics for $x$ and
$T_{l,r}$ up to times for which the mechanical equilibrium is
reached. Actually for such times the dynamics is essentially given by
the zeroth order equations (\ref{eq:vxlin}-\ref{eq:Trlin}) which are
the same for both the original and modified equations: this explain
their behavior in the mechanical regime. As one can see the main
difference in the dynamics is that retaining the terms $\langle
(V-v_x)v\rangle_{l,r}^{(1)}$ leads the system to stay for a longer
time interval in a state of approximate mechanical equilibrium and
thus to a slower relaxation to equilibrium. The evolution of the
piston temperature $M{\sigma_V^2}$ shown in Fig.~\ref{fig:relax2}
(bottom) clearly show the difference in the two dynamics, and in
particular the breaking of equipartition at equilibrium for the
modified dynamics.

It should be stressed that the behaviors shown in the above figures is not
related to the peculiar parameters choice as it has been verified in other
simulations (not shown).

The picture emerging from this comparison is that the model we
introduced goes to the correct equilibrium state and respects the
basic physics principles as conservation of energy and its
equipartition. On the other hand, the relaxation is slower than the
one of the microscopic model it should correspond to (i.e. the
randomized gas). As shown in the above figures the terms responsible
for this slowing down are the ones coming from the fluctuation of the
piston velocity in the expansion of the function $G$. At present we
dot not have a definite understanding either of this difference in the
dynamics of the macroscopic equations and of the randomized gas, or of
the very good agreement between the modified equations and the
randomized gas. We suspect that the contribution to the relaxation
time of these terms might be counterbalanced by the resummation of
higher order terms in the $\msuM$ expansion.  However, carrying the
analysis to higher order terms is very demanding due to the
proliferation of terms in the expansion, and therefore we shall not
discuss it here.

We conclude this section by mentioning that the equations derived by
Gruber and coworkers~\cite{lesne} predict a reasonable relaxation time
and equipartition at the same time.  We remark that going to higher
order in their approach requires to include into the description also
higher order moments of the piston velocity, namely
$\overline{V^3},\dots$; while within our approach the description
remains at the level of the second order moment because the statistics
is constrained to remain Gaussian (Maxwell-Boltzmann) for the gases
and the piston. This may be the origin of the difference between our
and their derivation.  Their equations at the first order might likely
correspond to a resummation of higher order terms in our
approach, meaning that their closure with the second moment of the
velocity is exact while ours remains only approximate.

\section{Conclusions}
\label{sec:3}

In the framework of kinetic theory, we derived a set of deterministic
equations describing the evolution of the macroscopic variables in the
adiabatic piston problem. Our basic assumptions are that at each time
the gases in the two compartments are perfect, spatially homogeneous,
and described by the Maxwell-Boltzmann statistics. Thus, at the level
of simulations, a (randomized) gas model has been introduced with aim
to have microscopic model respecting such assumptions.  We obtained a
set of five ordinary differential equations for the variables that
describe the macroscopic state of the system, namely the mean position
of the piston, the average velocity of the piston, the temperatures of
the gases in the two compartments and the second moment of the piston
velocity. The equations are derived up to the first order in $\msuM$. 

At the zeroth order they describe a deterministic piston characterized
by a velocity distribution collapsed on the mean, namely
$P(V)=\delta(V-v_x)$. This is enough to solve the problem of finding
the final state of mechanical equilibrium and the result coincides
with that derived in Ref.~\cite{lesne1} by using a different approach.

At the first order the fluctuations of the piston velocity, now assumed
to be Gaussian, allow for recovering the correct final thermodynamic
equilibrium. Although the evolution of the macroscopic observables
provided by this set of equations is in good qualitative agreement
with simulations of the randomized gas, we found some quantitative
discrepancy for the relaxation timescales. 

Apart from the performance in comparing with the simulations, we would
like to stress the conceptual aspects of the method we developed. It
allows for a transparent description of the macroscopic dynamics of a
nontrivial non-equilibrium problem, similarly to how the perfect gas
law can be derived from the microscopic collisions by using elementary
kinetic theory.

\begin{acknowledgements}
We thanks E.~Caglioti, F.~Cecconi, B.~Crosignani, P.~Di Porto,
A.~Lesne, G.P.~Morriss, and C.~Van den Broeck for useful discussions,
remarks and correspondence.  This work has been partially supported by
PRIN2003 ``Complex systems and many body problems'', and PRIN2005
``Statistical mechanics of complex systems'' by MIUR.  S.P. has
benefited from a MEC-MIUS joint program (Italy-Spain Integral
Actions).
\end{acknowledgements}


\begin{thebibliography}{200}

\bibitem{callen}H.B. Callen,
 {\it Thermodynamics}
(J. Wiley ans Sons, New York, 1963).

\bibitem{feynman}R.P. Feynman,
{\it Lectures in Physics I}
(Addison-Wisley, New York, 1965).

\bibitem{gruberlesne}C. Gruber and A. Lesne,
in {\it  Encyclopedia of Mathematical Physics}
 J.P. Francoise, G. Naber, T.S. Tsun (eds.) (Elsevier, 2006).


\bibitem{Ruchardt} E. R\"uchardt, 
  Z. Phys.~\textbf{30}, 58 (1929).

\bibitem{exp1} O.L. de Lange and J. Pierrus, 
  Phys. Rev. E~\textbf{57}, 5520 (1998).

\bibitem{exp2} J. Pierrus and O.L. de Lange, 
Phys. Rev. E~\textbf{56}, 2841 (1997).

\bibitem{curzon} A.E. Curzon, 
Am. J. Phys.~\textbf{37}, 404 (1969).


\bibitem{lieb}E. Lieb,
Physica A~\textbf{263}, 491 (1999).


\bibitem{bustamante} C. Bustamante, J. Liphardt and F. Ritort, 
Phys. Today~\textbf{58}, 43 (2005).

\bibitem{crosi1} B. Crosignani and P. Di Porto,
Europhys. Lett.~\textbf{53}, 290 (2001).

\bibitem{vandenbroeck1} C. Van den Broeck, 
New J. Phys.~\textbf{7}, 10 (2005).

\bibitem{crosignani} B. Crosignani, P. Di Porto and M. Segev,
Am. J.  Phys.~\textbf{64}, 610 (1996).

\bibitem{gruber1} C. Gruber, Eur. J. Phys.~\textbf{20}, 259 (1999). 

\bibitem{gruber2} C. Gruber and J. Piasecki, Physica A~\textbf{268},
  258 (1999).

\bibitem{gruber3} C. Gruber and L. Frachebourg, Physica
  A~\textbf{272}, 392 (1999).

\bibitem{morrisgruber02} G.P. Morriss and C. Gruber, 
  J. Stat. Phys.~\textbf{109}, 549 (2002).

\bibitem{gruberpache02} C. Gruber, S. Pache, Physica A~\textbf{314},
345 (2002).

\bibitem{lesne1} C. Gruber, S. Pache and A. Lesne,
  J. Stat. Phys.~\textbf{108}, 669 (2002).

\bibitem{lesne} C. Gruber, S. Pache and A. Lesne,
  J. Stat. Phys.~\textbf{112}, 1177 (2003).


\bibitem{morrisgruber03} G.P. Morriss and C. Gruber  , 
  J. Stat. Phys.~\textbf{113}, 297 (2003).

\bibitem{Mansour} M.M. Mansour, C. Van den Broeck, and E. Kestemont,
  Europhys. Lett.~\textbf{69}, 510 (2005); M.M. Mansour,
  A.L. Garcia, and F. Baras, Phys. Rev. E~\textbf{73}, 016121
  (2006).

\bibitem{sinai} N. Chernov, J. L. Lebowitz, and Ya. Sinai,
  J. Stat. Phys.~\textbf{109},  529 (2002).

\bibitem{lebo1} N. Chernov and J.L. Lebowitz, 
J. Stat. Phys.~\textbf{109}, 507 (2002).

\bibitem{Caglioti} E. Caglioti, N. Chernov and J.L~ Lebowitz,
Nonlinearity~\textbf{17}, 897 (2004).


\bibitem{vanderbroek2} E. Kestemont, C. Van den Broeck and M. Malek Mansour,
Europhys. Lett.~\textbf{49}, 143 (2000).

\bibitem{epaps} See EPAPS associated to this paper.

\bibitem{HOO} W.G. Hoover, {\it Time reversibility, computer
  simulations, and chaos}, pg.~47 (World Scientific, Singapore, 2001).
  
\end{thebibliography}
\end{document}